\documentclass{aa}  

\usepackage{graphicx}
\usepackage{txfonts}
\usepackage{xspace}
\usepackage{natbib}
\usepackage{color}
\usepackage{textcomp}
\newcommand{\Msun}{$M_{\odot}$\xspace}
\newcommand{\Lsun}{$L_{\odot}$\xspace}
\newcommand{\Msunyr}{$M_{\odot}$\,yr$^{-1}$\xspace}

\newcommand{\kms}{km\,s$^{-1}$\xspace}

\newcommand{\COthree}{$^{12}\rm{CO}\,(3-2)$\xspace}
\xspace
\xspace
\xspace
\newcommand{\micron}{$\mu$m}
\newcommand\crush{\textsc{crush}\xspace}
\newcommand\Mdotdustpd{$\dot{M}_{\rm{pd,d}}$}
\newcommand\Mdotdust{$\dot{M}_{\rm{d}}$}
\newcommand\Mshell{$M_{\rm{sh,d}}$}
\newcommand\Rshell{$R_{\rm{sh}}$}
\newcommand\dRshell{$\Delta R_{\rm{sh}}$}
\newcommand{\vdust}{$\varv_{\rm{exp,d}}$\xspace}
\newcommand{\agrain}{$a_{\rm{d}}$\xspace}
\begin{document}

   \title{Properties of dust in the detached shells around U~Ant, DR~Ser, and V644~Sco}


  \author{M.~Maercker
          \inst{1}
          \and
          T.~Khouri \inst{1}
          \and
          E.~De Beck \inst{1}
          \and
          M.~Brunner \inst{2}
          \and
          M.~Mecina \inst{2}
          \and
          O.~Jaldehag \inst{1}
          }

     \institute{Department of Space, Earth and Environment, Chalmers University of Technology, Onsala Space Observatory, 43992 Onsala, Sweden\\
   \email{maercker@chalmers.se}
         \and
             Department of Astrophysics, University of Vienna, T\"urkenschanzstr. 17, 1180 Vienna, Austria\\
             }

  \date{Received 18 June 2018 / Accepted 17 July 2018}

 
  \abstract
   {Asymptotic giant branch (AGB) stars experience strong mass loss driven by dust particles formed in the upper atmospheres. The dust is released into the interstellar medium, and replenishes galaxies with synthesised material from the star. The dust grains further act as seeds for continued dust growth in the diffuse medium of galaxies. As such, understanding the properties of dust produced during the asymptotic giant branch phase of stellar evolution is important for understanding the evolution of stars and galaxies. Recent observations of the carbon AGB star R~Scl have shown that observations at far-infrared and submillimetre wavelengths can effectively constrain the grain sizes in the shell, while the total mass depends on the structure of the grains (solid vs. hollow or fluffy).}
   {We aim to constrain the properties of the dust observed in the submillimetre in the detached shells around the three carbon AGB stars U~Ant, DR~Ser, and V644~Sco, and to investigate the constraints on the dust masses and grain sizes provided by far-infrared and submm observations.} 
   {We observed the carbon AGB stars U~Ant, DR~Ser, and V644~Sco at 870\,\micron~using LABOCA on APEX. Combined with observations from the optical to far-infrared, we produced dust radiative transfer models of the spectral energy distributions (SEDs) with contributions from the stars, present-day mass-loss and detached shells. We assume spherical, solid dust grains, and test the effect of different total dust masses and grain sizes on the SED, and attempted to consistently reproduce the SEDs from the optical to the submm.}
   {We derive dust masses in the shells of a few 10$^{-5}$\,\Msun. The best-fit grain radii are comparatively large, and indicate the presence of grains between 0.1\,\micron--2\,\micron. The LABOCA observations suffer from contamination from \COthree, and hence gives fluxes that are higher than the predicted dust emission at submm wavelengths. We investigate the effect on the best-fitting models by assuming different degrees of contamination and show that far-infrared and submillimetre observations are important to constrain the dust mass and grain sizes in the shells.}
   {Spatially resolved observations of the detached shells in the far-infrared and submillimetre effectively constrain the temperatures in the shells, and hence the grain sizes. The dust mass is also constrained by the observations, but additional observations are needed to constrain the structure of the grains. }


   \maketitle
%

\section{Introduction}

Cosmic dust is a key ingredient in star and planet formation and is critical in the evolution of galaxy properties~\citep{forestinico1997,herwigco2004,schneideretal2014,mancinietal2015}. Understanding the origins and properties of dust throughout the history of the universe is important for understanding the evolution of the universe itself. However, the source of dust in the early universe remains unclear, and even in the local universe the dominant origin of cosmic dust is unknown and the detailed properties of the dust grains are very uncertain.

Generally there are two main production sites of cosmic dust: dust formed in the winds of asymptotic giant branch (AGB) stars and dust formed in the ejecta of supernova (SN) explosions. In the interstellar medium (ISM), the expelled particles then act as seed particles for further growth and reprocessing~\cite[e.g.][]{dwek1998}. In the local universe (that is the Milky Way and nearby galaxies), it is likely that the dominant producer of interstellar dust are AGB stars~\citep[for example contributing up to 70\% of the dust in the Large Magellanic Cloud;][]{schneideretal2014}. While SNe can be significant producers of dust~\citep[e.g.][]{matsuuraetal2015}, a large fraction of the dust may get destroyed in the reverse shock~\citep{schneideretal2014,bocchioetal2016}. The balance between dust from AGB stars and SNe additionally depends on the assumed initial mass function (IMF).

In the early universe, the situation is even more complicated. Stars that will evolve to become AGB stars have main sequence masses of 0.8-8\,\Msun. The lifetime of these stars of in the order of a few billion years, making them less likely to contribute significantly, or at all, to the production of dust at high redshifts, as they will not have had sufficient time to evolve to the AGB stage. However, dusty galaxies are observed also in the early universe. Recently, ALMA observations revealed a dusty galaxy at redshift 7.5~\citep{watsonetal2015}. At lower redshifts (z$\sim4-5$), AGB stars may be able to account for the amount of observed dust in starburst galaxies~\citep[e.g.][]{michalowskietal2010}. While understanding the origin and evolution of the dust is important for understand the evolution of galaxies throughout the universe~\citep[e.g.][]{bekki2015}, the reprocessing that grains undergo in the ISM complicates the determination of the origin of the observed dust.

For AGB stars, the dust grains likely play a dominant role in the stellar mass-loss process itself, since radiation pressure of stellar light on the grains is thought to regulate the mass-loss rate. An accurate description of the mass-loss process, and the dust that is tightly linked to it, is essential for correct models of stellar evolution, which estimate stellar yields.~\citep[e.g.][]{woitke2006,hofnerco2007}.

Dust properties and yields from AGB stars have predominantly been determined through modelling of the spectral energy distribution (SEDs) of galactic AGB stars, complemented with spectral observations of various dust features~\citep[e.g.,][]{groenwegen2012,blommaertetal2014,rauetal2017}, typically from UV wavelengths out to the far-infrared (FIR). The type of dust around AGB stars is roughly separated into two regimes, with carbon-rich dust forming around carbon AGB stars (with atmospheric carbon-to-oxygen ratios C/O$>$1), while M-type AGB stars (with atmospheric C/O$<$1) form silicate grains. In the case of carbon AGB stars, the dust particles are dominated by amorphous carbon grains, with possible contributions from MgS and/or SiC~\citep[e.g.][]{honyco2004}. In models, the grains are generally assumed to be of sub-micron size (0.01\,\micron~to 0.5\,\micron), spherical, and solid~\citep[e.g.][]{schoieretal2005}. The optical properties of the grains are determined through laboratory measurements under various conditions~\citep[e.g.][]{rouleauco1991,preibischetal1993,zubkoetal1996,jageretal1998,suh2000}. 

The composition of dust grains around stars is commonly studied using the infrared excess produced from thermal dust emission. In particular, the resonance features of the different dust species at infrared wavelengths is a useful tool for determining the composition of circumstellar dust. Studying dust at submm wavelengths is problematic for several reasons. The uncertainties in temperature and density profiles, and dust properties in the submm, combined with the very low spatial resolution of single-dish telescopes, make it difficult to unambiguously constrain the detailed dust content around AGB stars. Detached shells around carbon AGB stars provide ideal laboratories where these problems can be overcome. These shells are likely created during the high-mass-loss-rate phases of a thermal pulse (TP), and retain their shape while expanding away from the star due to interaction with a previous, slower wind~\citep{olofssonetal1996,steffenco2000,mattssonetal2007}. Detached shells of CO and dust have been detected around seven carbon AGB stars~\citep{schoieretal2005}, and have been studied in optical scattered light and molecular line emission, probing both the dust and gas~\citep[e.g.][]{olofssonetal1993a,olofssonetal1996,olofssonetal2010,delgadoetal2001,delgadoetal2003a,maerckeretal2010,maerckeretal2012,maerckeretal2014,maerckeretal2016}. The shells have angular sizes on the sky of $\approx$10\arcsec--60\arcsec, typically have widths of only a few arcseconds, and are remarkably spherically symmetric. Owing to their simple geometry, they make it possible to study the circumstellar environment with a well-defined density distribution, and largely avoiding line of sight confusion. They may also be the only way to study the changes a star experiences during and after a TP directly, constraining an important process during AGB evolution.

Recently, \cite{brunneretal2018} presented observations towards the carbon AGB star R~Scl at 870\,\micron. The star is surrounded by a detached shell of dust and gas at $\approx$20\arcsec\,from the star, likely to have been formed during a TP $\approx$2000 years ago. Combining the submm observations with data from optical to FIR wavelengths, they determine the effect of different grain properties on the model SED. The observations at wavengths $>$100\,\micron\,strongly constrain the dust in the shell to be dominantly in $\approx$0.1\,\micron\,sized grains. The total estimated dust mass in the shell is most strongly affected by the assumed structure of the grains (solid vs. hollow and/or fluffy). The FIR and submm observations hence effectively constrain the grain size in the shell and, provided the structure of the grains can be determined, also the total dust mass.

Here we present observations at submm wavelengths towards the carbon AGB stars U~Ant, DR~Ser, and V644~Sco, and modelling of the observed SEDs with a focus on the FIR and submm properties and the constraints these may set on the estimated dust masses in the shells and the grain sizes. All three sources are surrounded by detached shells of gas and dust~\citep{olofssonetal1996,delgadoetal2001,delgadoetal2003a,ramstedtetal2011,maerckeretal2014}. In Sect.~\ref{s:observations} we describe the basic source parameters and observations. In Sect.~\ref{s:modelling} we describe the modelling strategy. In Sects.~\ref{s:results} and~\ref{s:discussion} we present and discuss the results. We end the paper with concluding remarks in Sect.~\ref{s:conclusions}.

\section{Observations}
\label{s:observations}

\subsection{Sources}
\label{s:sources}

We have analysed observations towards the carbon-rich AGB stars U~Ant, DR~Ser, and V644~Sco. All sources are irregular variables and have been shown to be surrounded by detached shells of dust and gas~\citep{olofssonetal1996,delgadoetal2001,delgadoetal2003a,ramstedtetal2011,maerckeretal2014}. The shells are geometrically thin (\dRshell/\Rshell$\approx$0.15 -- 0.2) and remarkably spherical. They were likely created during recent thermal pulses, and have ages between 1300 -- 2800 years, assuming a constant expansion velocity since creation of the shell~\citep{schoieretal2005}. The parameters of the sources are summarised in Table~\ref{t:stparams}.

While DR~Ser and V644~Sco only show one shell, U~Ant has at least two detached shells: shell 3 and shell 4~\citep[at \Rshell=43\arcsec and \Rshell=50\arcsec, respectively, following the naming convention in earlier publications; ][]{delgadoetal2001,delgadoetal2003a,maerckeretal2010}. Based on optical observations of polarised light scattered by dust grains and in the resonance lines of Na and K, \cite{maerckeretal2010} concluded that shell 4 is dominated by dust, while shell 3 is dominated by gas. High-resolution images of CO emission observed with ALMA towards U~Ant show that the shell of gas coincides with shell 3~\citep{kerschbaumetal2017}. When the density in the shell decreases as the shell evolves, larger grains will eventually separate from the gas and accelerate away due to additional radiation pressure from the star, forming the outer shell 4. Shell 3 may possibly retain a small amount of dust in small grains. The thermal emission observed in PACS images at 70\,\micron~and 160\,\micron~appears to only come from shell 3~\citep{kerschbaumetal2010}, while \cite{arimatsuetal2011} also detect a double-shell with a dominant mass component in shell 4 using AKARI observations between 65\,\micron~--160\,\micron. They confirm the scenario in which large grains have separated from the gas, creating shell 4, while smaller grains are retained in shell 3. 

In this paper we will assume that all the dust in the shells around U~Ant is located in shell 4. In Sect.~\ref{s:results} we discuss the effect on the models of adding a shell of small grains at the position of shell 3. 

\begin{table*}
\centering
\caption{Stellar parameters. The luminosities $L$ and present-day dust-mass-loss rates (\Mdotdustpd) are assumed and are consistent with modelled upper limits~\citep{schoieretal2005}. Effective temperatures $T_{\rm{eff}}$ and distances $D$ are derived in the dust modelling. The dust-expansion velocities (\vdust)  are assumed to be the same as the gas-expansion velocities, and are taken from~\cite{schoieretal2005}. The assumed shell radii $R_{\rm{sh}}$ and full-width-half-maximum (FWHM) $\Delta R_{\rm{sh}}$ are also given~\citep[from][]{ramstedtetal2011,maerckeretal2010,maerckeretal2014}.}
\label{t:stparams}
\begin{tabular}{lccccccc}
\hline
\hline
Source 	& $L$ 	& $T_{\rm{eff}}$ 	& $D$ 	& \Mdotdustpd 	& \vdust & \Rshell & \dRshell\\
		& [\Lsun]	&	[K]				& [pc]	& [\Msunyr] & [\kms] & ["] & ["]\\
\hline
U~Ant	& 4000		& 2300			& 230	&	1$\times10^{-9}$	& 4 & 50 & 7.0\\
DR~Ser	& 4000		& 2550			& 720	& 	1$\times10^{-9}$	&5 & 7.6& 1.2\\
V644~Sco& 4000		& 2315			& 760	&   1$\times10^{-9}$	&5 & 9.4& 2.0\\
\hline
\end{tabular}
\end{table*}

\subsection{Spectral energy distributions}
\label{s:seds}

We constructed SEDs for U~Ant, DR~Ser, and V644~Sco using archived data from the optical to the submm. \emph{JHKLM} photometry is the same as the data used for SED modelling in~\cite{schoieretal2005}. We additionally used AKARI and IRAS photometry extracted from the respective archives. For U~Ant we added Herschel/PACS~\citep{kerschbaumetal2010} and Herschel/SPIRE observations. See Sect.~\ref{s:spireuant} for the treatment of the SPIRE data. Finally, we added new data at 870\,\micron~for all sources (Sect.~\ref{s:labocaobs}). All observations used for the SEDs are summarised in Table~\ref{t:observations}.

\subsection{LABOCA observations at 870\,\micron}
\label{s:labocaobs}

For all sources, we obtained continuum observations at 870\,\micron~(345 GHz) using the Large Apex BOlometer CAmera \citep[LABOCA;][]{labocaref}, a 295-channel bolometer array mounted on the Atacama Pathfinder EXperiment~\citep[APEX\footnote{This publication is based on data acquired with the Atacama Pathfinder Experiment (APEX) under programme ID 094.F-9313. APEX is a collaboration between the Max-Planck-Institut f\"ur Radioastronomie, the European Southern Observatory, and the Onsala Space Observatory.};][]{apexref} telescope in Chile.  The observations were carried out with 223 live pixels. Pointing corrections were well within 5\arcsec\/ in azimuth and 4\arcsec\/ in elevation, whereas the full width at half maximum (FWHM) of the APEX beam at 870\,\micron\/ is 19.2\arcsec. The focus was stable to within 0.2\,mm in all three spatial axes x, y, and z. Corrections for atmospheric attenuation were obtained from opacity measurements in regular skydips, performed about every $1-2$ hours. The absolute flux calibration is based on observations of a primary (Mars) and multiple secondary calibrators (e.g., HL~Tau, CW~Leo) throughout the observing sessions. All data were obtained under very dry and stable weather conditions with precipitable water vapour (pwv) values of $0.4-1.2$\,mm.

\begin{figure}
\centering
\includegraphics[width=8cm]{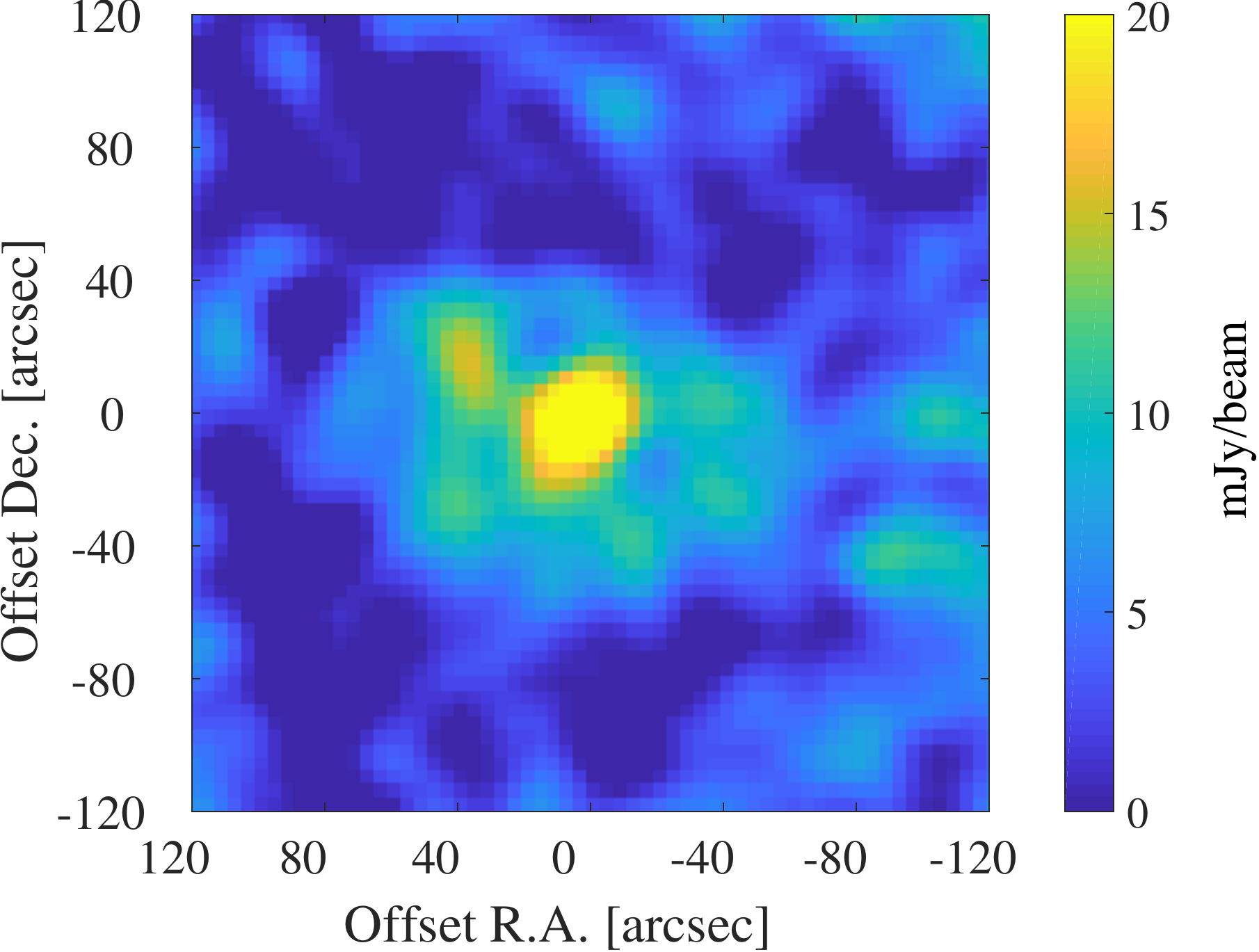}
\caption{LABOCA observations of U~Ant at 870\,\micron. The image is smoothed with a Gaussian kernel of 8\arcsec. The colourscale is in mJy/beam of the smoothed image.}
\label{f:uantlaboca}
\end{figure}

Given the FWHM beam of LABOCA, we performed single pointing observations on DR~Ser and V644~Sco where the shell diameters are smaller than the beam, and mapping observations of U~Ant. Observations of DR~Ser and V644~Sco were obtained on 10 and 11 August 2014 in the wobbler on-off mode, with a 1\,Hz wobbler frequency in a symmetric nodding pattern with a wobbler amplitude of 25\arcsec, and 30\,seconds-long nod phases. LABOCA's bolometer channel 71, the most stable and most sensitive bolometer, was used as the reference channel.  For U~Ant, we obtained an on-the-fly (OTF) map 
on 11, 12, and 13 August 2014 and 26, 27, 28 June and 8 July 2015, with a final map size of $27.7\arcmin\times20.9\arcmin$ (Fig.~\ref{f:uantlaboca}).

The data inspection and reduction was carried out with \crush\footnote{\texttt{http://www.submm.caltech.edu/$\sim$sharc/crush/}} release 2.15-1, a reduction and imaging tool developed for bolometer arrays \citep{kovacs2006,kovacs2008_crush}. Optimising the reduction procedure for faint point sources, we obtain a flux density of $13.3\pm2.3$\,mJy\,beam$^{-1}$ for DR~Ser, and of $39.0\pm3.9$\,mJy\,beam$^{-1}$ for V644~Sco, after 145\,minutes and 48\,minutes of integration time, respectively. U~Ant shows a detached shell of dust located at $\approx50$\arcsec\/ with a width of 7\arcsec~\citep{maerckeretal2010}, and the shell is resolved in the LABOCA observations. Therefore we optimised the data reduction for U~Ant for extended emission. The total integration time in the final map was 607 minutes. Smoothing of 3\farcs76 was applied, resulting in an effective FWHM of the beam of 19\farcs86. The total flux was measured to be $210\pm 50$\,mJy\,beam$^{-1}$ in a circular region of radius $65$\arcsec\/. The background flux was estimated within a ring of $100\arcsec<R_{\rm{bg}}<140$\arcsec.  The measured fluxes are given in Table~\ref{t:observations}.

\subsection{Contamination by \COthree line emission}
\label{s:cocont}

Observations with LABOCA measure the flux at 345\,GHz within a 60\,GHz window. Consequently it also covers emission from the \COthree line at 345.795\,GHz, which may dominate the detected flux. We roughly estimate the amount of flux from \COthree contributing to the LABOCA observations using single-dish data and radiative transfer models.

We model the \COthree emission based on the results by~\cite{schoieretal2005} and the observations summarised there. We do not attempt to model the line in detail, but to get a good enough fit to the line in order to get a rough estimate of the total flux from the shell. We use a 1-dimensional, non-LTE radiative transfer code based on the monte-carlo method~\citep{schoierco2001}. For each source we model the \COthree emission using the shell gas-mass, shell temperature, and expansion velocity derived by ~\cite{schoieretal2005}, as well as the parameters for the present-day mass-loss. The density distribution of the shells is a step-function centred at radius \Rshell with a width \dRshell. The density distribution across the shell is calculated assuming a constant mass-loss rate at constant expansion velocity for a period that corresponds to the width of the shells. The mass-loss rate is chosen so that the total mass in the shell is consistent with the results in~\cite{schoieretal2005}. The models are constrained by single-dish observations obtained with SEST (for U~Ant and V644~Sco) and JCMT (for DR~Ser), with FWHM beams of 14\arcsec. The observations resolve the shells (which have diameters larger than the beam, see Table~\ref{t:stparams}), and hence do not measure the entire \COthree emission. We therefore estimate the total contribution to the LABOCA fluxes by calculating the emission from the models in beams that cover the entire shells. In this case we get total \COthree fluxes of 0.06\,Jy, 0.007\,Jy, and 0.015\,Jy for U~Ant, DR~Ser, and V644~Sco, respectively, corresponding to 38\%, 70\%, and 38\% of the observed LABOCA fluxes.

It is difficult to determine the degree of contamination without observations that map the \COthree emission from the entire shell and present-day mass-loss. A non-homogeneous distribution in the shells and detailed beam-shapes make the estimates based on the radiative transfer models uncertain. However, our rough estimates here show that a significant fraction of the observed LABOCA flux may come from \COthree line emission instead of continuum dust emission. Due to the uncertainty in determining the degree of contamination, the best-fit models of the shells presented here do not include the LABOCA observations. However, we investigate the constraints set by submm observations by assuming varying degrees of \COthree contamination, and emphasise the importance of accurate, and spatially resolved, measurements in the FIR and submm (Sect~\ref{s:submmconst}).

\subsection{SPIRE observations of U~Ant}
\label{s:spireuant}

U~Ant was observed with SPIRE onboard Herschel as part of the MESS guaranteed time programme (PI: M. Groenwegen). The data were first presented in~\cite{groenewegenetal2011}. The flux values given there are extracted from apertures of 25\arcsec in radius, hence missing most of the emission from the shell. We therefore retrieved the level 2 maps at 250\,\micron, 350\,\micron, and 500\,\micron\/ from the archive using the final calibration. The data were calibrated assuming an extended source. The beam of SPIRE at 250\,\micron, 350\,\micron, and 500\,\micron\/ is 18\farcs1, 25\farcs2, and 36\farcs6, respectively. The flux was extracted from the maps within a circular aperture with a radius of 80\arcsec. The background was subtracted measured in a ring between $100\arcsec<R_{\rm{bg}}<140$\arcsec.



\begin{table*}
\centering
\caption{Observed data for U~Ant, DR~Ser, and V644~Sco. $F_{\lambda}$ and $\Delta F_{\lambda}$ are the flux density and flux density error (1$\sigma$) at wavelength $\lambda$, respectively. For V644~Sco, the values of $\Delta F_{\lambda}$ for some wavelengths are unavailable and are assumed to be 10\% of $F_{\lambda}$. The origin of the data is indicated in the last column. }
\label{t:observations}
\begin{tabular}{cccccccc}
\hline
\hline
			& \multicolumn{2}{c}{U~Ant} &\multicolumn{2}{c}{DR~Ser} &\multicolumn{2}{c}{V644 Sco} & Reference\\
$\lambda$ 	& $F_{\lambda}$& $\Delta F_{\lambda}$ &$F_{\lambda}$& $\Delta F_{\lambda}$ &$F_{\lambda}$& $\Delta F_{\lambda}$ & \\
$\mu$m		& \multicolumn{2}{c}{[Jy]}&\multicolumn{2}{c}{[Jy]}&\multicolumn{2}{c}{[Jy]}& \\
\hline
0.44        & 1.12		& 0.58      & 0.08		& 0.08     	& --     	&  --      	& \cite{kerschbaumco1999} \\   
1.24        & 597.60	& 22.02     & 71.20		& 1.71     	& 38.70		& 3.87   	& ''                    \\
1.63        & 1127.52	& 41.54     & 85.22		& 3.14    	& --     	&  --      	& ''                    \\
1.66        & --      	&  --       & --        &  --       & 85.90		& 8.59   	& ''                    \\
2.16        & --      	&  --       & --        &   --      & 103.00	& 10.30  	& ''                    \\
2.19        & 1167.55	& 43.01     & 98.64		& 3.63     	& --     	& --       	& ''                    \\
3.79        & 749.41	& 27.61     & 70.78		& 2.61     	& --     	& --       	& ''                    \\
4.29        & --      	&  --       & --        &   --      & 65.92		& 6.59   	& ''                    \\
4.35        & --      	&  --       & --        &   --      & 73.00		& 7.30   	& ''                    \\
4.64        & 319.03	& 23.51     & --		& --     	& --     	& --    	& ''                    \\
8.28        & --      	&  --       & 21.67		& 2.10      & 24.67		& 2.47   	& ''                    \\
12.13       & --      	&  --       & 11.60		& 1.10      & 15.03		& 1.50   	& ''                    \\
14.65       & --      	&  --       & 7.73		& 0.80      & 8.90		& 0.89   	& ''                    \\
21.30       & --      	&  --       & 4.74		& 0.40      & 6.41		& 0.64   	& ''                    \\ \rule{0pt}{2.9ex}
8.610       & 263.5    	&  14.6     & 21.00		& 0.50      & 22.71    	&  0.15     & AKARI (archive)     \\
18.40       & 61.5     	&  2.3      & 6.16		& 0.13      & --     	&  --       & ''                    \\
65.00       & 25.8$^*$    	& 5.3       & 4.40		& 0.30      & --     	&  --       & ''                    \\
90.00       & 20.1$^*$    	& 4.2       & 3.68		& 0.51      & --     	&  --       & ''                    \\
140.00      & 8.4$^*$   	& 3.1       & --		& --   	    & --     	&  --       & ''                    \\
160.0       & 3.5$^*$   	& 2.2       & 1.28		& 0.19      & --     	&   --      & ''                    \\ \rule{0pt}{2.9ex}
12.00       & 167.50	& 12.23     & 15.95		& 1.16     	& 17.62		& 1.76   	& IRAS (archive)        \\
25.00       & 44.81		& 4.03      & 6.66		& 0.60     	& 8.99		& 0.90   	& ''                    \\
60.00       & 27.11		& 3.25      & 5.960		& 0.71     	& 19.30		& 1.93   	& ''                    \\
100.00      & 21.14		& 2.60      & --        & --       	& --     	& --    	& ''                    \\ \rule{0pt}{2.9ex}
70.00		& 27.1		& 4.1		& --		& --		& --		& --		& PACS~\citep{kerschbaumetal2010}					\\
160.00		& 7.4		& 1.2		& --		& --		& --		& --		& ''					\\ \rule{0pt}{2.9ex}
250.00      & 1.89		& 0.28      & --        & --      	& --     	& --      	& SPIRE (archive, Sect.~\ref{s:spireuant})                 \\
350.00      & 0.73		& 0.11      & --        & --       	& --     	& --      	& ''                    \\
500.00      & 0.25		& 0.04      & --        & --      	& --     	& --      	& ''                    \\ \rule{0pt}{2.9ex}
870.00      & 0.16		& 0.02      & 0.01		& 0.002    	& 0.04		& 0.004  	& APEX LABOCA (this paper, Sect.~\ref{s:labocaobs})           \\
\hline
\multicolumn{8}{l}{$^*$ \tiny{For U~Ant the AKARI data from 65\,\micron~to 160\,\micron~was taken from~\cite{arimatsuetal2011} instead of from the AKARI archive.}}
\end{tabular}
\end{table*}

\section{Modelling}
\label{s:modelling}

The SEDs were modelled using the Monte Carlo dust radiative transfer code MCMax~\citep{minetal2009}. The code calculates the dust radiative transfer including absorption, re-emission and scattering processes using the Monte Carlo method. It uses an input radiation field (in our case the central star), the dust density profile (the present-day wind and detached shell), and the optical dust properties, and produces output SEDs and images that can be directly compared to the observations. We initially modeled the star and present-day wind without the shell (Sect.~\ref{s:modstpd}). The resulting SED was then used as an input for the models of the shells (see Sect.~\ref{s:dcsemods}). 

\subsection{Star and present-day wind}
\label{s:modstpd}

At wavelengths shorter than $\approx12$\,\micron, the SEDs will be dominated by the stellar radiation and present-day wind. In a first step, we constrained the parameters of the star. For all sources we assumed a stellar luminosity of 4000\Lsun (see below), and a present-day dust mass-loss rate \Mdotdustpd$=10^{-9}$\,\Msunyr. By fitting the radiative transfer models to the observed SED points at $\lambda<$12\,\micron, we constrained the distances and effective temperatures. The derived parameters are given in Table~\ref{t:stparams}. We note that the derived true stellar effective temperature depends on how the models include the warm dust close to the star, and may be slightly higher.

Assuming 4000\Lsun~for all sources introduced an uncertainty in the estimated distances. For Mira variables and semi-regular variables it is possible to derive a luminosity using a period-luminosity (PL) relationship~\cite[e.g.][]{knappetal2003}, and hence derive more accurate distances. However, in this case all sources are irregular variables, and the PL-relationships can not be applied. Hipparcos (and recently Gaia) parallaxes are intrinsically uncertain for AGB stars because of their sizes (typically the same or larger than the measured parallaxes), and possible variable features across the stellar discs~\citep{khourietal2016a,vlemmingsetal2017}. For typical luminosities on the AGB (a 2.5\Msun star will have a luminosity of 2000 -- 6000 \Lsun during the majority of the thermally pulsing AGB), we estimated the uncertainty in the derived distances to be less than 20\%.

The dust-density radial profile of the present-day wind was calculated assuming a homogeneous, constant wind expanding at a constant velocity. In this case the density profile is proportional to the present-day dust mass-loss rate (\Mdotdustpd) and the dust-expansion velocity (\vdust):

\begin{equation}
\label{e:pddensprof}
\rho_{\rm{d}}(r) \propto\frac{\dot{M}_{\rm{pd,d}}}{r^2\varv_{\rm{exp,d}}}.
 \end{equation} 

\noindent 
It is difficult to unambiguously determine \vdust, and hence the real parameter that is constrained by the models is the ratio \Mdotdustpd$/\varv_{\rm{exp,d}}$. Assuming full coupling between the dust and expanding gas in the present-day wind (that is equal velocities for the dust and gas), it is possible to constrain $\varv_{\rm{exp,d}}$ using molecular line observations of the stellar wind, and hence estimate \Mdotdustpd. Here we assumed dust velocities of \vdust=4~\kms~for U~Ant~\citep{kerschbaumetal2017}, {\vdust=5}~\kms~for DR~Ser and V644~Sco~\citep{schoieretal2005} based on measured gas-expansion velocities, and one constant grain size of \agrain=0.1\,\micron. The value of \Mdotdustpd=$10^{-9}$\Msunyr is an upper limit. Higher mass-loss rates over-predict the NIR observations. For the study here the exact value of the present-day mass-loss is not important. The primary objective is that the radiation field from the star and present-day mass-loss are reproduced accurately as input for the detached-shell models (Sect.~\ref{s:dcsemods}).

\subsection{Detached shells}
\label{s:dcsemods}

MCMax produces an output SED for the star and present-day wind that was used as an input spectrum for the models of the detached shells. The shells were assumed to have a gaussian density distribution with a radius \Rshell~and a full-width-half-maximum (FWHM) \dRshell~(see Table~\ref{t:stparams}). Such a density distribution is consistent with the dust-scattered light observations~\citep{delgadoetal2001,delgadoetal2003a,maerckeretal2010,maerckeretal2014,olofssonetal2010}. For each source we calculated a base-model of the SED including the star and present-day mass-loss, and assuming a shell mass of \Mshell=$1\times10^{-5}$\Msun, typical for estimated dust masses in detached shells~\citep{schoieretal2005,maerckeretal2010,olofssonetal2010}. The resulting SED contains emission from the star, present-day wind, and the detached shell. Subtracting the star and present-day wind contributions from the total SED gives the SED for the shell in the base-model only. In order to determine the emission from shells with different masses, the base-model can be scaled accordingly, and then added back to the emission from the star and present-day wind. For the densities considered here, the shell is optically thin at all wavelengths.  

For all sources we varied the shell-masses from $0.2\times10^{-5}$\Msun to $20\times10^{-5}$\Msun in steps of $0.01\times10^{-5}$\Msun. The best-fit model was determined by minimizing the $\chi^2_{red}$

\begin{equation}
\label{e:redchi2}
\chi^2_{red}=\frac{1}{N_{\rm{obs}}-1}\sum{n=1}\frac{(F_{\rm{mod}}-F_{\rm{obs}})^2}{\Delta F_{\rm{obs}}^2},
\end{equation}

\noindent
where $F_{\rm{mod}}$ and $F_{\rm{obs}}$ are the modelled and observed fluxes, respectively, $\Delta F_{\rm{obs}}$ is the uncertainty in the observed flux, $N_{\rm{obs}}$ is the number of observations used in the fit, and the sum goes over all observations included in the fit. The shells only contribute to the SEDs at wavelengths longer than 12\,\micron, and we only fited to observations at $\lambda$>12\,\micron.

\subsection{Grain properties}
\label{s:grainprops}

The contribution to the SED from the dust grains depends on the temperature of the grains. Since the shells have a well-defined geometry, with all the dust at essentially the same distance from the star, the temperature for any particular type of dust is fixed. A change in the emission can only be achieved by changing the grain properties. This was investigated in detail for the detached shell around the carbon AGB star R~Scl~\citep{brunneretal2018}. They modelled the effect on the SED by changing the assumed dust opacities, grain sizes, composition, and structure (solid vs. hollow sphere, and fluffy grains). While the different properties affected the total estimated dust mass in the shell, the only parameter that significantly affected the temperature of the grains, and hence the shape of the SED, was the grain size. Since in this paper we are primarily interested in the constraints the observations set on the grain sizes, this is the only grain property we varied.

\begin{figure}[t]
\centering
\includegraphics[width=8cm]{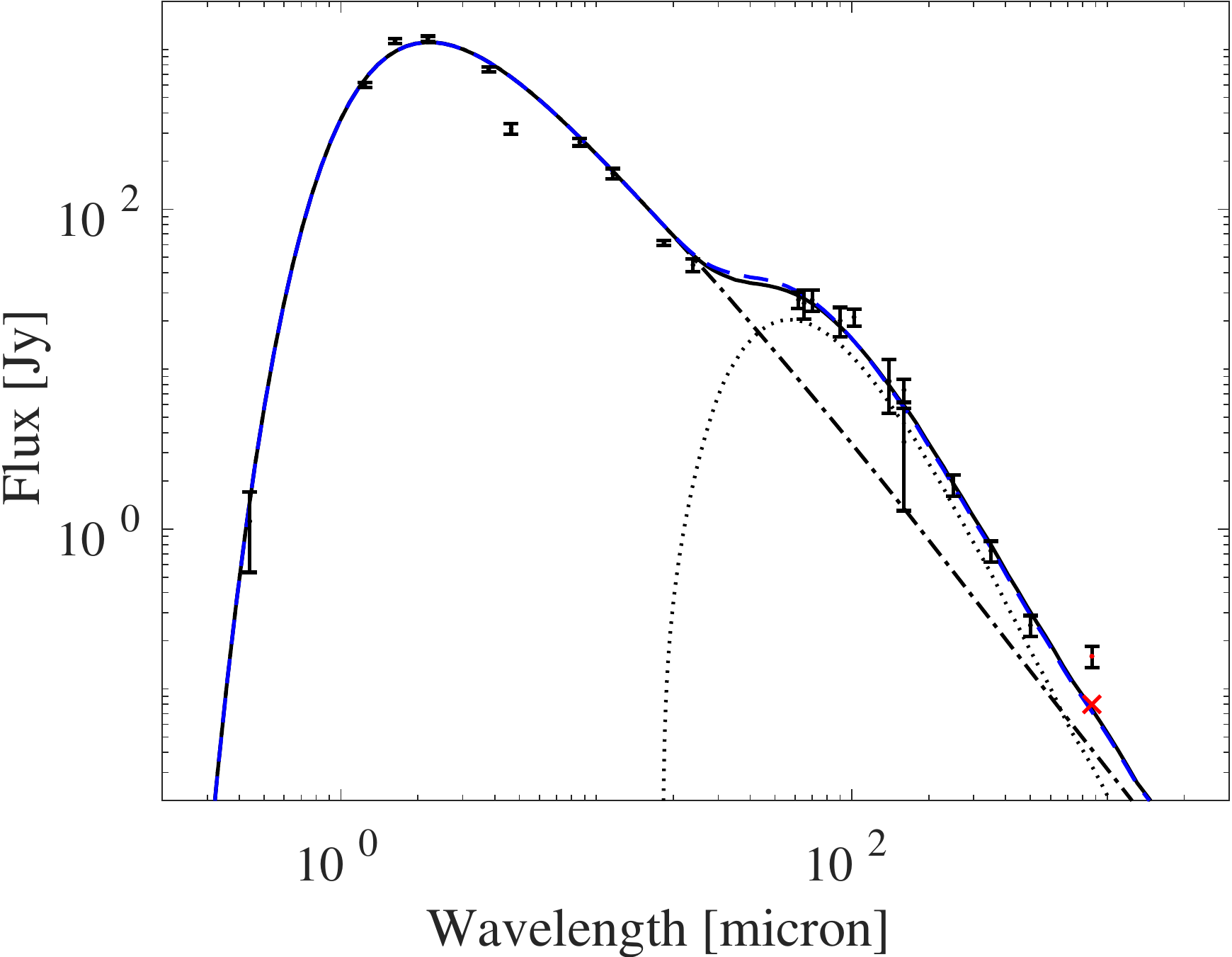}
\includegraphics[width=8cm]{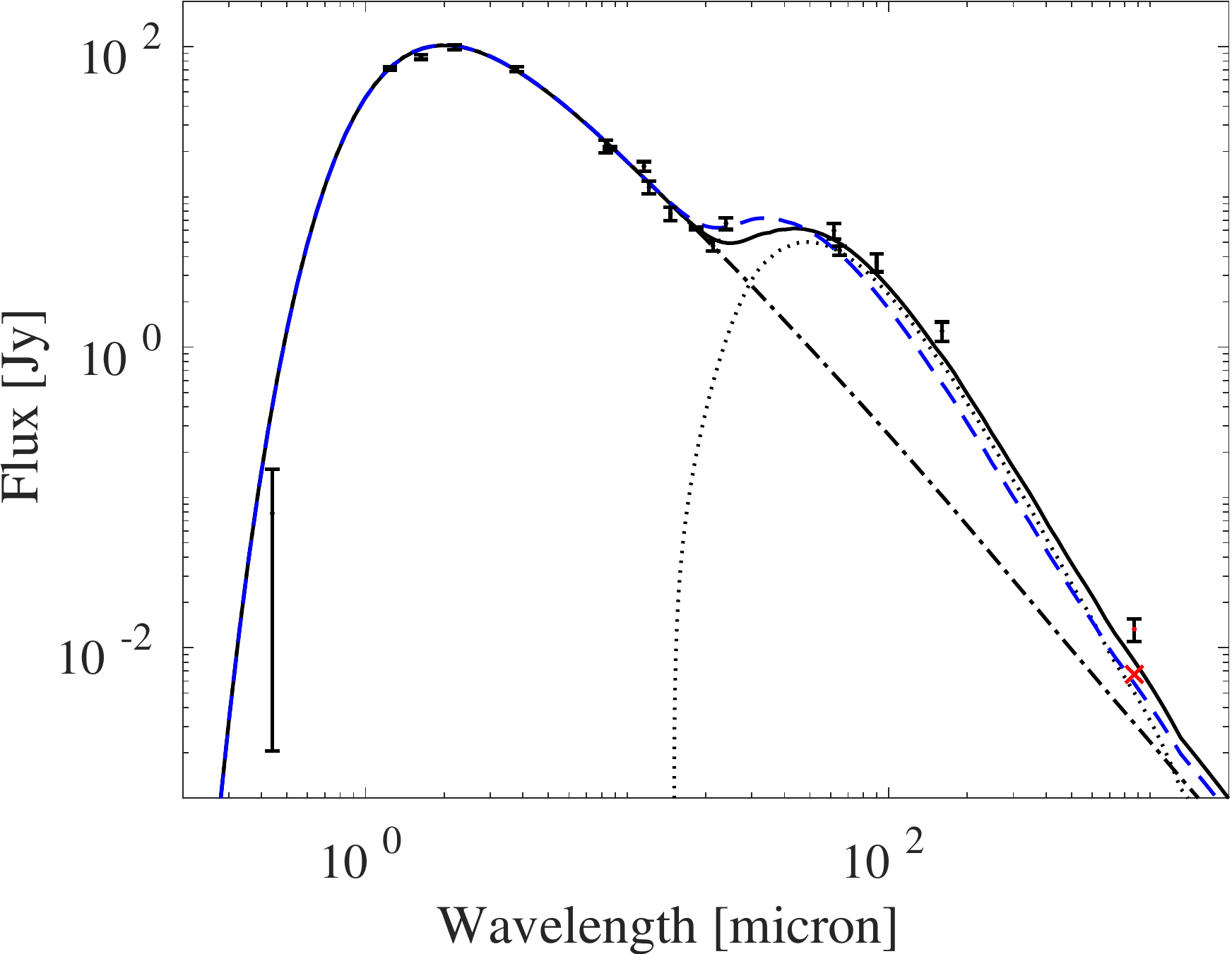}
\includegraphics[width=8cm]{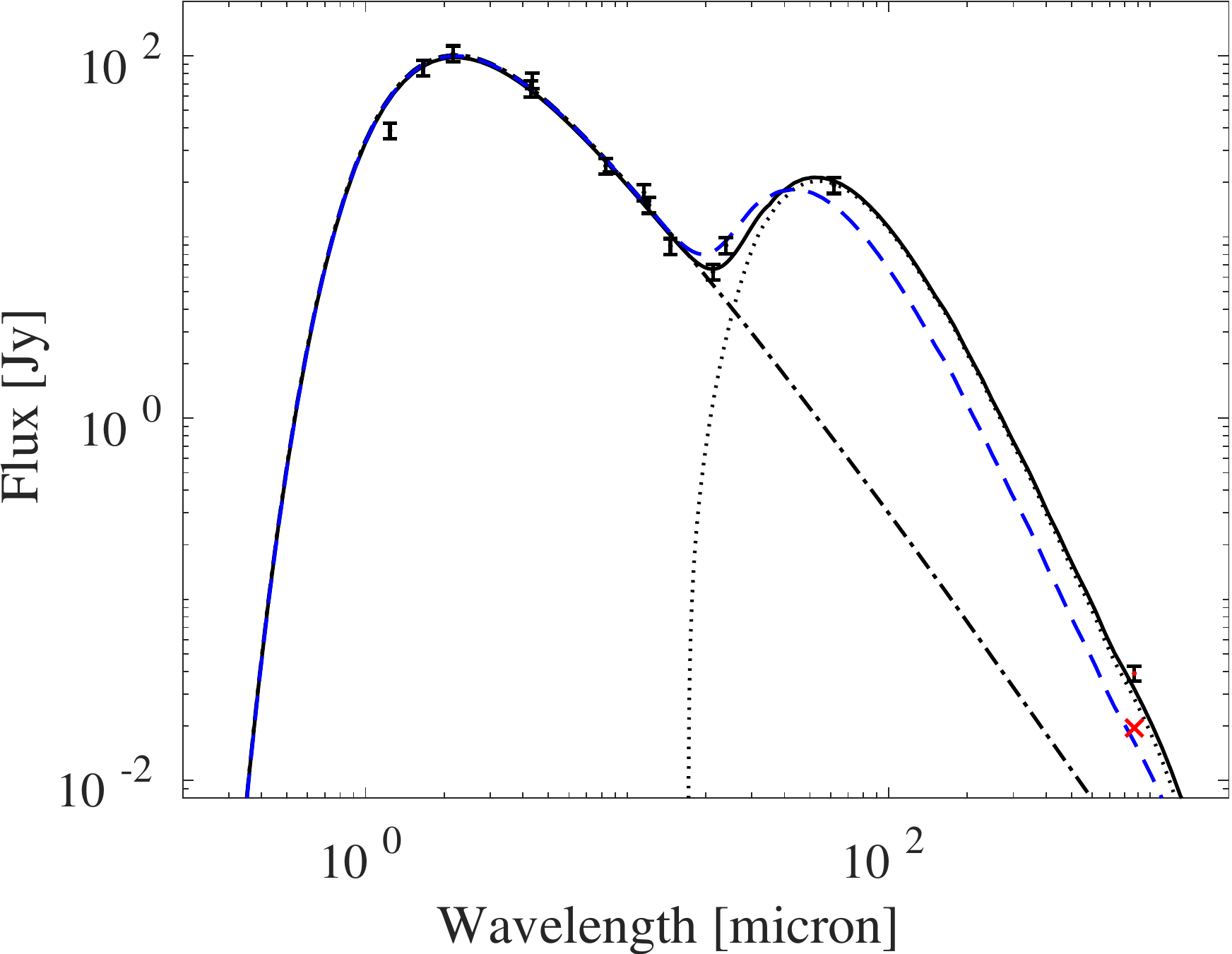}
\caption{SEDs of U~Ant (top), DR~Ser (middle), and V644~Sco (bottom) with best-fit models for 0.1\,\micron~sized grains (blue-dashed lines), and 1.0\,\micron~(U~Ant) and 2.0\,\micron~(DR~Ser and V644~Sco) sized grains (solid lines). The dot-dashed lines show the contribution to the total SED from the star and present-day wind, the dotted lines the contribution to the SED from the shells. The red crosses indicate the 50\%-level of the LABOCA fluxes.}
\label{f:allseds}
\end{figure}

For all models, we used opacities for amorphous carbon grains from~\cite{suh2000}, assuming solid spheres. For the present-day wind we assumed a constant grain size of 0.1\,\micron. For the shells we calculated different models assuming constant, single grain-sizes in the shells of \agrain=0.1\,\micron, 0.5\,\micron, 1.0\,\micron, 2.0\,\micron, and 5.0\,\micron. The $\chi^2_{red}$ was calculated for each \agrain and \Mshell, and we determine the 1$\sigma$ uncertainties in the dust mass. 

\section{Results}
\label{s:results}

The results of the radiative transfer models of the shells are presented in Table~\ref{t:shresults}. We present the best-fit models assuming the same grains as in the present-day wind (\agrain=0.1\,\micron), and the best-fit models when treating the grain size as a free parameter. The best-fit models are shown in Fig.~\ref{f:allseds}.

\begin{figure}
\centering
\includegraphics[width=8cm]{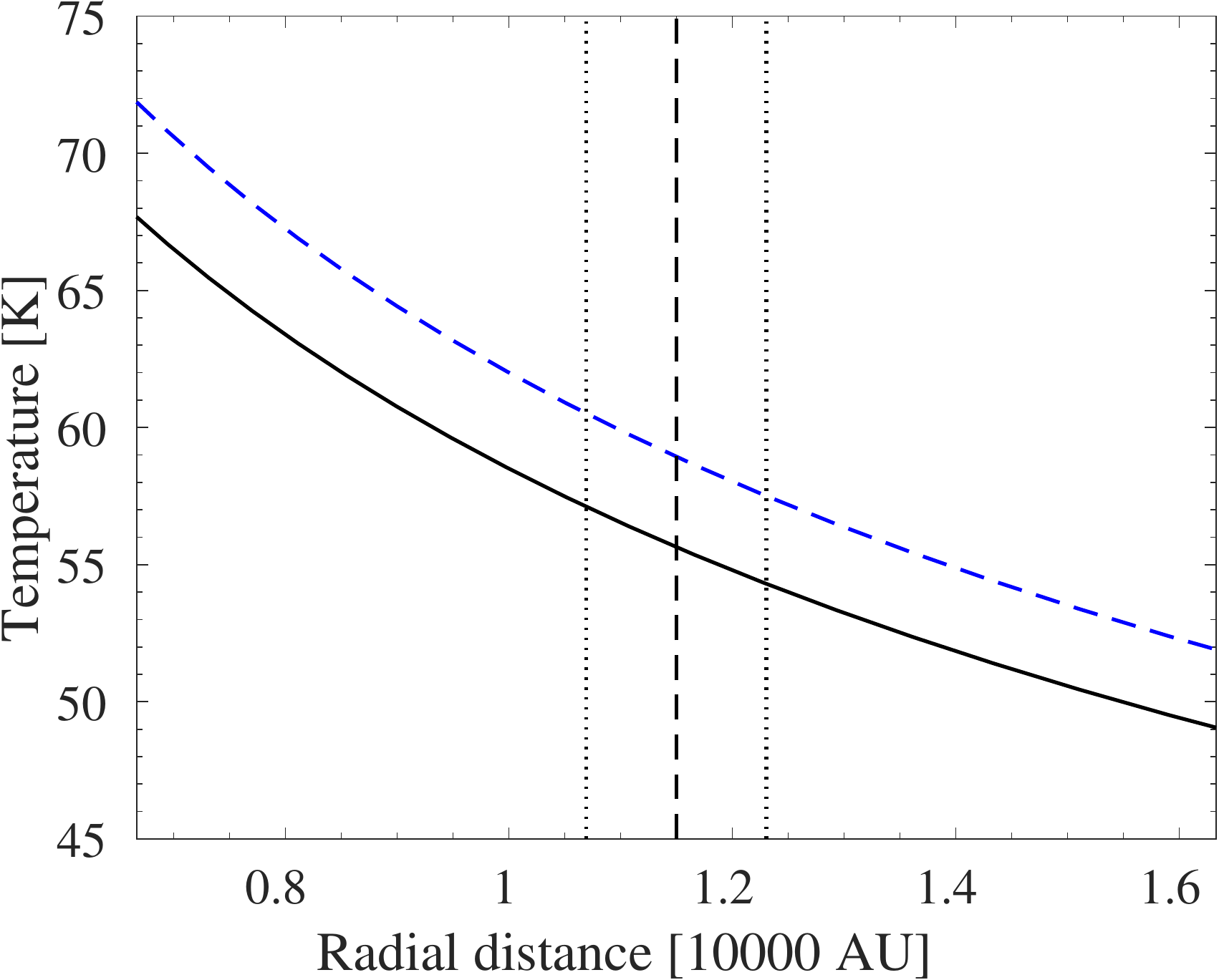}
\includegraphics[width=8cm]{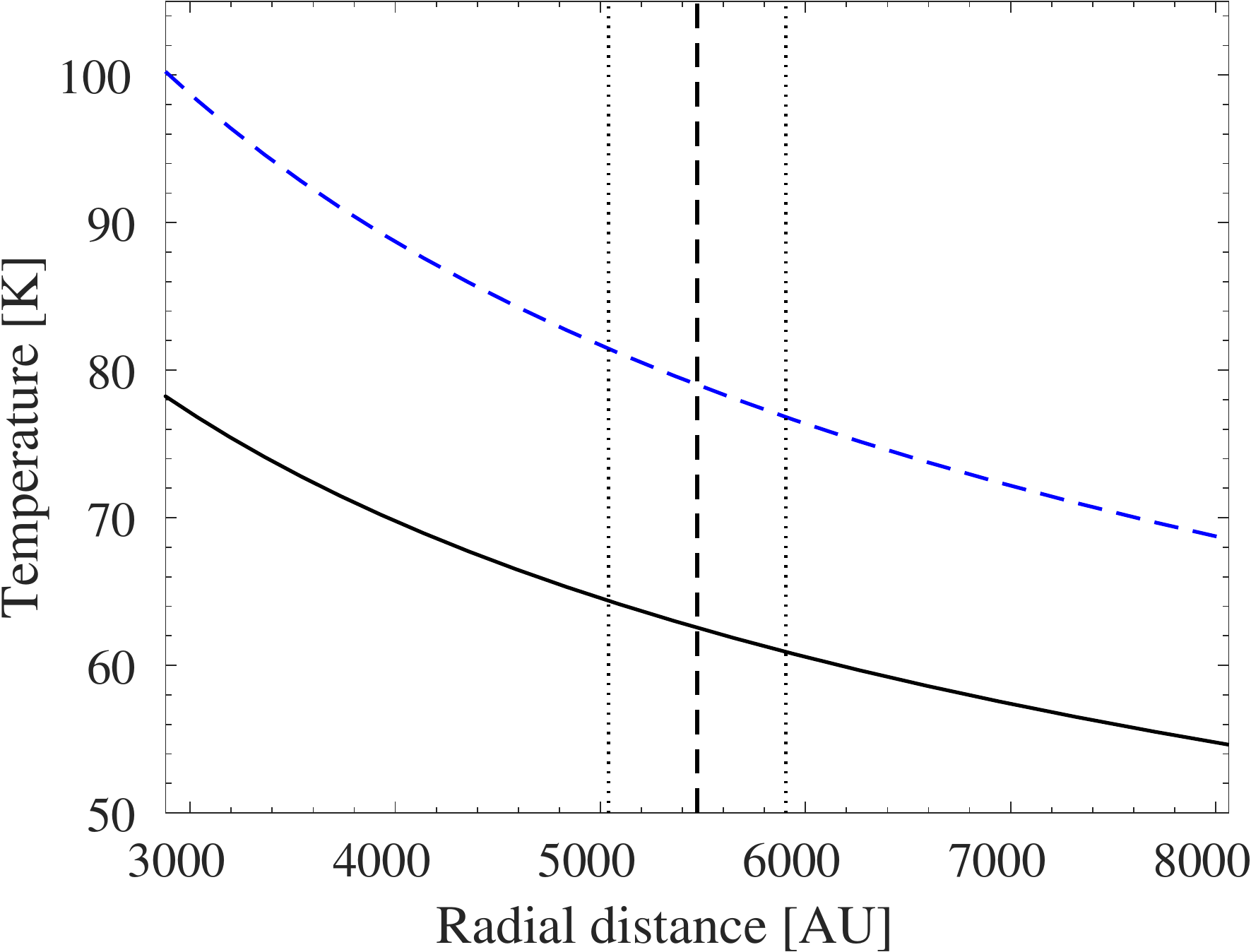}
\includegraphics[width=8cm]{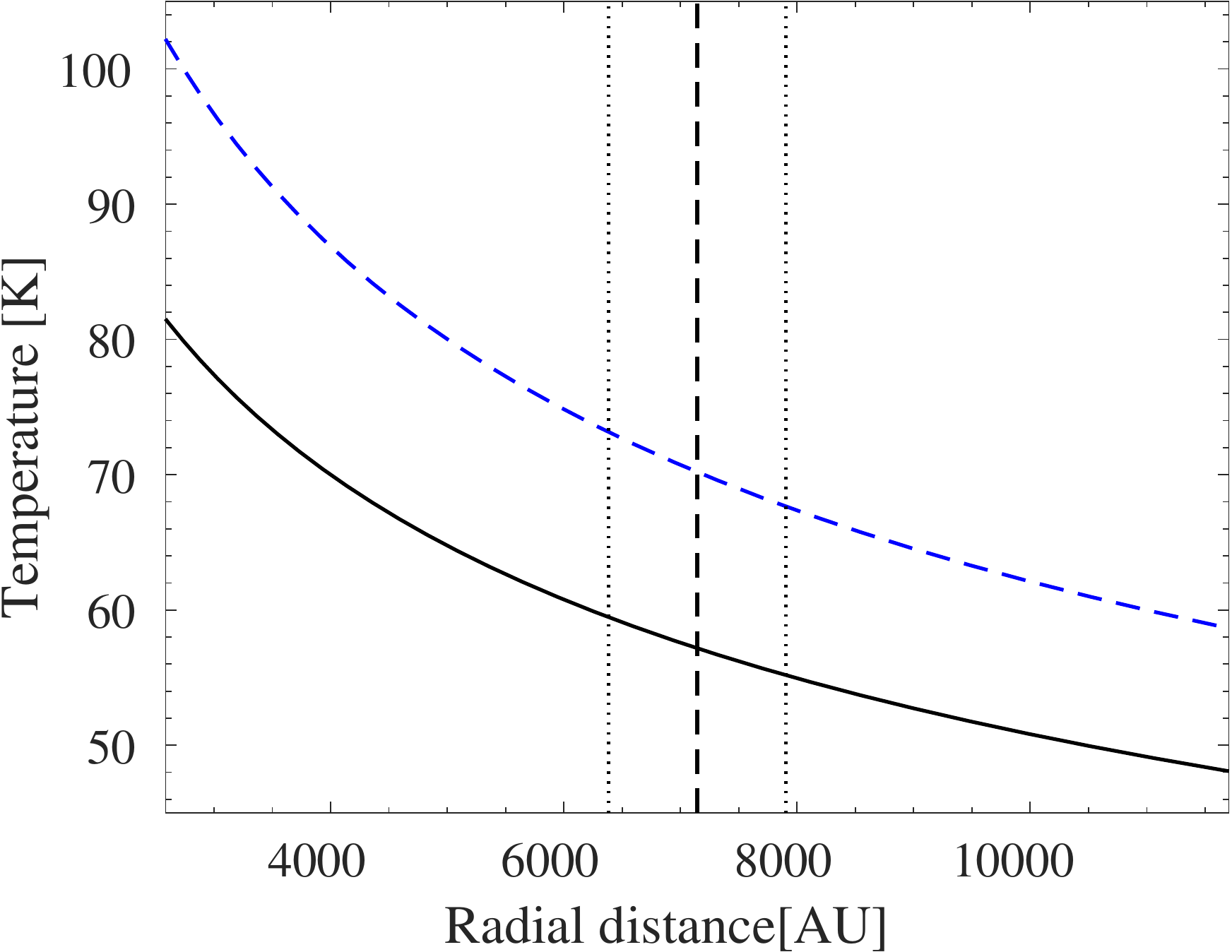}
\caption{Temperature profiles at the positions of the shell for U~Ant (top), DR~Ser (middle), and V644~Sco (bottom). The vertical dashed lines show the radii of the shells, the vertical dotted lines indicate the FWHM of the shells. The blue dashed line shows the temperature profile for models with \agrain=0.1\,\micron, and black solid line shows the temperature profile for models with \agrain=1.0\,\micron~(U~Ant) and \agrain=2.0\,\micron~(DR~Ser and V644~Sco).}
\label{f:tprofs}
\end{figure}

For 0.1\,\micron-sized grains we derive dust masses in the shells of (1.6$\pm$0.4)$\times10^{-5}$\Msun, (1.0$\pm$0.2)$\times10^{-5}$\Msun, and (6.0$\pm$1.2)$\times10^{-5}$\Msun for U~Ant, DR~Ser, and V644~Sco, respectively. Treating the grain size as a free parameter, we derive shell dust-masses of (1.9$\pm$0.4)$\times10^{-5}$\Msun, (2.4$\pm$0.4)$\times10^{-5}$\Msun, and (17.0$\pm$3.2)$\times10^{-5}$\Msun, respectively. The best-fit grain-sizes in these cases are 1.0\,\micron~for U~Ant, and 2.0\,\micron~for DR~Ser and V644~Sco. The corresponding temperature profiles at the position of the shells are shown in Fig.~\ref{f:tprofs}. As expected, larger grains result in a colder shell. Using grains with \agrain=0.1\,\micron, \cite{schoieretal2005} derive shell dust-masses of (13$\pm$12)$\times10^{-5}$\Msun, (3.5$\pm$2.5)$\times10^{-5}$\Msun, and (14$\pm$9)$\times10^{-5}$\Msun for U~Ant, DR~Ser, and V644~Sco, respectively. These fits only consider observations up to 100\,\micron. However, the SED from the detached shells peaks just below 100\,\micron, and the shell mass is mainly constrained at longer wavelengths. Adding data at longer wavelengths therefore constrains the models significantly better. Additionally, while~\cite{schoieretal2005} manage to get good fits to the observed SEDs, the shell sizes they derive are significantly larger than shown by observations (using their distances and shell radii the apparent sizes of the shells on the sky would have to be 130\arcsec, 14\arcsec, and 16\arcsec for U~Ant, DR~Ser, and V644~Sco, respectively). Adding the spatial constraints for the shells, models using 0.1\,\micron-sized grains result in too high dust temperatures, and hence a worse fit to the data, and explains the difference in our results compared to~\cite{schoieretal2005} for the 0.1\,\micron-sized grains. The combination of spatial constraints and observations at long wavelengths hence effectively constrains the grain sizes and masses in the shells.

\begin{figure}
\centering
\includegraphics[width=8cm]{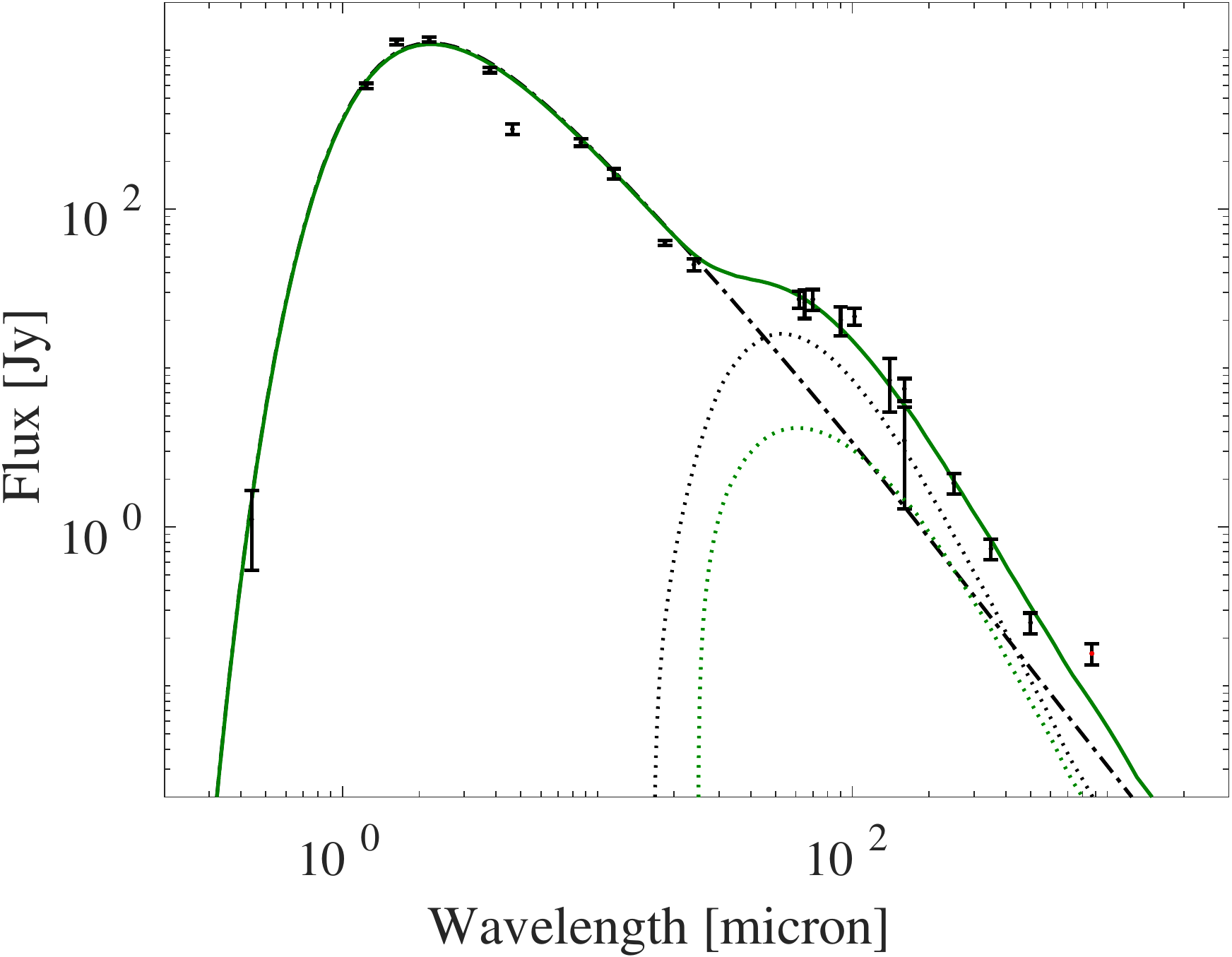}
\caption{SED of U~Ant with a shell of dust added at the position of shell 3. The solid green line gives the total SED including star and present-day wind (dot-dashed line), and shells 3 (green dotted) and 4 (black dotted). }
\label{f:shell3fig}
\end{figure}

For U~Ant we additionally investigate the possible contribution from dust in shell 3. \cite{arimatsuetal2011} derive a total dust mass of \Mshell=$1.6\times10^{-5}$\Msun in shell 4, in excellent agreement with our results. For shell 3, they derive a dust mass of \Mshell=$1.9\times10^{-7}$\Msun, that is two orders of magnitude lower. Assuming that shell 3 contains small dust grains that are retained by the gas, we add a dust shell at the position of shell 3 with \agrain=0.01\,\micron~and a mass \Mshell=$2\times10^{-7}$\Msun. The best-fit models in this case are essentially the same as for models with only shell 4. Increasing the mass in shell 3 by a factor of 10 to \Mshell=$2\times10^{-6}$\Msun, results in lower \Mshell~for shell~4 by $\approx$43\% (to \Mshell=$1.06\times10^{-5}$\Msun), and worse fits to the SED (with a $\chi^2_{red}=1.6$; Fig.~\ref{f:shell3fig}). In particular, the best-fit grain-size for shell 4 in this case is \agrain=0.5\,\micron. The increased emission from shell 3 increases the emission also at FIR wavelengths. As a consequence, smaller grains are forced into shell 4, increasing the temperature in this shell and decreasing the FIR emission to still fit the observations. A further increase of the mass in shell 3 results in increasingly bad fits to the data. Hence, we can conclude that the mass in shell 3 is at least one order of magnitude lower than for shell 4, and is dominated by small grains, in line with previous results~\citep{maerckeretal2010,arimatsuetal2011}.

\begin{table}
\centering
\caption{Results of the radiative transfer modelling of the shells. The LABOCA data are not taken into account in the fitting.}
\label{t:shresults}
\begin{tabular}{lcccc}
\hline
\hline

Source 	& \agrain	& \Mshell	& $\Delta$\Mshell	& $\chi^2_{red}$\\
		& [\micron]	&	\multicolumn{2}{c}{[$10^{-5}$\Msun]}& \\ 
\hline
U~Ant	& 0.1	& 1.6	& 0.4	&  1.4\\
		& 0.5	& 1.5	& 0.4	&  1.5\\
		& \bf{1.0}	& \bf{1.9}	& \bf{0.4}	&  \bf{1.3}\\
		& 2.0	& 2.6	& 0.6	&  1.9\\
		& 5.0	& 4.2	& 1.1	&  4.6\\
DR~Ser	& 0.1	& 1.0	& 0.2	&  8.4\\
		& 0.5	& 0.9	& 0.2	& 9.2\\
		& 1.0	& 1.4	& 0.2	&	4.8\\
		& \bf{2.0}	& \bf{2.4}	& \bf{0.4}	&  \bf{2.9}\\
		& 5.0	& 0.9	& 0.9	& 3.4\\
V644~Sco& 0.1	& 6.0	& 1.2	&  4.5\\
		& 0.5	& 5.0	& 1.0	&	5.8\\
		& 1.0	& 8.6	& 1.6	&	2.4\\
		& \bf{2.0}	& \bf{17.0}	& \bf{3.2}	&  \bf{1.8}\\
		& 5.0	& 39.8	& 8.2	& 6.7\\

\hline
\end{tabular}
\end{table}

\section{Discussion}
\label{s:discussion}


\subsection{Shell dust masses}
\label{s:dustmasses}

We derive dust masses in the shells of the order of a few $10^{-5}$\Msun. Comparing the dust-shell masses with estimates of the gas-shell masses~\citep{schoieretal2005}, we derive gas-to-dust ratios of 100 or less (in the case of V644~Sco the ratio is as low as 15). These values are considerably lower than expected from carbon-rich AGB stars~\citep[e.g.][]{ramstedtetal2008}, and would imply a more efficient dust production during creation of the shells. However, the gas-to-dust ratios are very uncertain and likely to be underestimated. The dust models assume solid spheres, and the derived \Mshell-values are likely upper limits. As was shown in~\cite{brunneretal2018}, the total dust mass decreases when using, for example, a distribution of hollow spheres or fluffy grains. The estimates of the gas-masses in the shells assume a CO/H$_2=1\times10^{-3}$ ratio to derive the total mass from observations of CO rotational emission lines~\citep{schoieretal2005}. This value is generally assumed for the outflows of carbon-rich AGB stars. However, the shells will have been exposed to the interstellar radiation field, dissociating the CO, and hence leading to an underestimation of the total gas masses in the shells. Combined, these effects would drive the gas-to-dust ratios to higher values, in line with what is expected towards AGB winds.

\subsection{Grain sizes}
\label{s:grainsizes}

For all sources we also find an indication for larger grains in the shells than typically are assumed to form around AGB stars, which is not as easily explained. U~Ant has the best-constrained SED in the FIR. For this source, the best-fit models for \agrain=0.1\,\micron~and \agrain=1.0\,\micron~are almost equally good. R~Scl has a similarly well-sampled SED in the FIR, and also here models with \agrain=0.1\,\micron~ fit the observations well. For DR~Ser and V644~Sco larger grains give better fits. However, here the SEDs are not very well sampled, and the grain-size is only an indication for potentially larger grains in these sources. Generally, grains with \agrain=0.1--1.0\,\micron~would be comparatively large compared to what is usually assumed for AGB winds. However, we note that pre-solar grains collected from meteorites, and that likely originated in AGB stars, can have sizes of several microns~\citep[e.g.][]{bernatowiczetal1996,xuetal2016}. Large grains may be a consequence of continued grain-growth in the expanding shell, where densities remain significantly higher than in a regular wind following a r$^{-2}$ density distribution~\citep{mattssonetal2007}. It should be noted that in the case of U~Ant and R~Scl, the dust in the shells can not be material swept up from the ISM. The interaction of the stellar wind with the ISM can clearly be seen to lie outside the detached shell in both cases. In all cases the observations at FIR and submm wavelengths effectively constrain the temperature in the shells, and hence the grain sizes, independent of other grain properties (see also Sect.~\ref{s:submmconst}).

\subsection{Shell evolution}
\label{s:evolution}

The detached shells are believed to be formed during the high mass-loss rate phases during a thermal pulse, with an increase in the wind-expansion velocity leading to a wind-wind interaction, shaping the shells~\citep[e.g.][]{olofssonetal1996,steffenco2000,mattssonetal2007}. In principle, the study of detached shells should hence allow us to constrain the properties of the mass-loss during and after a thermal pulse. This was done in detail using CO observations with ALMA towards R~Scl~\citep{maerckeretal2012,maerckeretal2016}. Assuming 200 years for the high mass-loss rate period during a thermal pulse, the results here would imply a dust mass-loss rate \Mdotdust~during the thermal pulse of $\approx10^{-7}$\,\Msunyr. Keeping in mind the uncertainties in the derived \Mdotdustpd-values, this would imply a drop in the dust-mass-loss rate of two orders of magnitude over the course of 1000-2000 years. A sudden drop in gas- and dust-mass-loss rates is expected from models of thermal pulses, with a slow increase during the inter-pulse period back to pre-pulse values~\citep{karakasco2007}, leaving the shells essentially devoid of dust and gas. This is contrary to what is found for R~Scl, where both the analysis of CO emission lines and dust continuum emission from inside the shell indicate that the shell in fact is filled with gas and dust~\citep{maerckeretal2012,maerckeretal2016,hankinsetal2018}. So far this was interpreted as a slower decline in mass-loss rate after the pulse than predicted by models. However, CO observations with ALMA towards U~Ant~\citep{kerschbaumetal2017}, and the results presented here, appear more in line with stellar evolution models. It is not clear how to reconcile the different results. The evolution of the mass loss during and after a thermal pulse is critical for stellar evolution models. The mass loss limits the time the star spends on the AGB, and hence the number of thermal pulses and periods of element-synthesis the star can experience. The mass-loss-rate evolution hence directly affects the yields of elements to the ISM from AGB stars.

\subsection{Constraints set by FIR and submm observations}
\label{s:submmconst}

For all sources we plot the measured LABOCA flux, and indicate the flux when assuming that 50\% of the flux is due to \COthree contamination (Fig.~\ref{f:allseds}). For all sources this degree of contamination appears consistent with the dust models. The SED is best sampled for U~Ant, and the FIR and submm observations constrain the dominant grain sizes to be $<$2.0\,\micron. Including the uncorrected LABOCA flux in these models does not change the estimated dust mass, but only results in a slightly worse fit. For DR~Ser and V644~Sco there is an increasing lack of observations at FIR and submm wavelengths. Including the uncorrected LABOCA data in the fit for DR~Ser results in a best-fit model with a grain size of \agrain=5.0\,\micron, with a significantly higher dust mass in the shell (\Mshell=5.3\,\Msun). In the case of V644~Sco, the best-fit model nearly reproduces the uncorrected flux at 870\,\micron~for grains with \agrain=2.0\,\micron, and there is no significant change in the best-fit model when including the LABOCA flux in the fit. However, models with small grains fit the observations significantly worse.

If we instead include the LABOCA observations in the fits and assume a degree of contamination of 50\%, the models for U~Ant again do not change significantly compared to not including the LABOCA observations. For DR~Ser the corrected submm observations now also result in best-fit models with 2.0\,\micron\, grains and similar values to not including the LABOCA observations. For V644~Sco the best-fit model is now achieved with smaller grains (\agrain=1.0\,\micron) and a lower mass (\Mshell=8.5\,\Msun). 

We note that the dust mass is only constrained when making assumptions on the grain structure~\citep{brunneretal2018}, and additional observations are neccessary to constrain the total dust mass (see Sect.~\ref{s:moreobs}).

\subsection{The need for more observations}
\label{s:moreobs}

The results indicate that spatially and spectrally resolved observations of dust at FIR and submm wavelengths can constrain the sizes of dust grains in detached-shell sources. The spatial resolution constrains the distance of the grains from the central source, and hence the temperature of the grains. The spectral resolution allows to measure the un-contaminated continuum emission. This is particularly important at submm wavelengths, where AGB stars show significant emission from molecular lines.

Our results show that FIR and submm observations also constrain the total dust mass in the shells. However, this is dependent on the assumed structure of the grains (that is, the observations effectively only constrain the number of dust grains). Additional observations that constrain the grain structure, in combination with spatially resolved FIR and submm observations are required to effectively constrain the total dust mass. The polarisation properties of porous grains in the optical may offer a possibility to determine the grain structure~\citep{kirchschlageretal2014}.

These grains observed in the detached shells may be similar to the dust released by AGB stars to the ISM. The grain size has implications for the survival of grains when transported into the ISM, and for further evolution in the ISM, where the grains act as seed particles for grain growth. 

\section{Conclusions}
\label{s:conclusions}

We present models of the dust continuum emission in the FIR and submm of the carbon AGB stars U~Ant, DR~Ser, and V644~Sco. The sources are surrounded by detached shells of dust and gas which likely were formed during recent thermal pulses. We derive dust masses in the shells that are consistent with previous estimates, but with significantly lower uncertainties. For all sources there is an indication of comparatively large grains (\agrain=0.1\,\micron--1.0\,\micron). The derived masses suggest a rapid decline in dust mass-loss rate after the thermal pulse, in line with stellar evolution models. 

For all sources we show that FIR and submm observations are needed to effectively constrain the grain sizes in the detached shells. Considering their ages and distances from the central stars, the properties of the dust in detached shell sources may be similar to the dust released to the ISM by AGB stars. In this context, the detached shell sources are ideal objects to probe the dust properties around AGB stars and the dust-contribution of low-mass stars to galaxies. Their simple geometry removes uncertainties in the dust density distribution and temperature. In order to further constrain the dust masses and grain sizes, spatially resolved observations at FIR, submm, and millimetre wavelengths are necessary. Observations with the Atacama Compact Array (ACA) in bands 3, 6, and 7 (870\,\micron--3\,mm) would effectively constrain the flux in the submm while allowing to identify contaminating molecular emission lines. Observations between 100\,\micron--500\,\micron~constrain the shape of the SED in the FIR towards the 870\,\micron~observations, effectively constraining the temperature of the dust in the shells, and hence the grain sizes, as well as the total dust mass in the shells.

The results have implications for the dust production in AGB stars, and their contribution to the ISM in the Milky Way. Taken at face value, it appears that more dust is produced in AGB stars, with larger grains than generally assumed. However, systematic and spatially resolved observations of the dust emission in the submm towards AGB stars are necessary to derive the properties of dust produced in low-mass stars. Additionally, it is not clear whether the dust formed in detached shells is representative of the dust produced during the AGB in general, and hence a sample of sources including all types of AGB stars needs to be studied.

\begin{acknowledgements} M. Maercker acknowledges support from the Swedish Research Council under grant number 2016-03402. T.K.  acknowledges support from the Swedish Research Council.  E.D.B. acknowledges funding by the Swedish National Spaceboard. M.B. acknowledges funding through the uni:docs fellowship of the University of Vienna and funding by the Austrian Science Fund FWF under project number P23586. The authors would like to thank the referee for constructive comments that improved the quality of the paper.
\end{acknowledgements}

\bibliographystyle{aa} 
\bibliography{../../../bib/maercker}

\begin{thebibliography}{55}
\expandafter\ifx\csname natexlab\endcsname\relax\def\natexlab#1{#1}\fi

\bibitem[{{Arimatsu} {et~al.}(2011){Arimatsu}, {Izumiura}, {Ueta}, {Yamamura},
  \& {Onaka}}]{arimatsuetal2011}
{Arimatsu}, K., {Izumiura}, H., {Ueta}, T., {Yamamura}, I., \& {Onaka}, T.
  2011, ApJ, 729, L19

\bibitem[{{Bekki}(2015)}]{bekki2015}
{Bekki}, K. 2015, ApJ, 799, 166

\bibitem[{{Bernatowicz} {et~al.}(1996){Bernatowicz}, {Cowsik}, {Gibbons},
  {Lodders}, {Fegley}, {Amari}, \& {Lewis}}]{bernatowiczetal1996}
{Bernatowicz}, T.~J., {Cowsik}, R., {Gibbons}, P.~C., {et~al.} 1996, ApJ, 472,
  760

\bibitem[{{Blommaert} {et~al.}(2014){Blommaert}, {de Vries}, {Waters},
  {Waelkens}, {Min}, {Van Winckel}, {Molster}, {Decin}, {Groenewegen},
  {Barlow}, {Garc{\'{\i}}a-Lario}, {Kerschbaum}, {Posch}, ƒ~{Royer}, {Ueta},
  {Vandenbussche}, {Van de Steene}, \& {van Hoof}}]{blommaertetal2014}
{Blommaert}, J.~A.~D.~L., {de Vries}, B.~L., {Waters}, L.~B.~F.~M., {et~al.}
  2014, A\&A, 565, A109

\bibitem[{{Bocchio} {et~al.}(2016){Bocchio}, {Marassi}, {Schneider}, {Bianchi},
  {Limongi}, \& {Chieffi}}]{bocchioetal2016}
{Bocchio}, M., {Marassi}, S., {Schneider}, R., {et~al.} 2016, A\&A, 587, A157

\bibitem[{{Brunner} {et~al.}(2018){Brunner}, {Maercker}, {Mecina}, {Khouri}, \&
  {Kerschbaum}}]{brunneretal2018}
{Brunner}, M., {Maercker}, M., {Mecina}, M., {Khouri}, T., \& {Kerschbaum}, F.
  2018, A\&A, 614, A17

\bibitem[{{Dwek}(1998)}]{dwek1998}
{Dwek}, E. 1998, ApJ, 501, 643

\bibitem[{{Forestini} \& {Charbonnel}(1997)}]{forestinico1997}
{Forestini}, M. \& {Charbonnel}, C. 1997, A\&AS, 123, 241

\bibitem[{{Gonz{\'a}lez Delgado} {et~al.}(2001){Gonz{\'a}lez Delgado},
  {Olofsson}, {Schwarz}, {Eriksson}, \& {Gustafsson}}]{delgadoetal2001}
{Gonz{\'a}lez Delgado}, D., {Olofsson}, H., {Schwarz}, H.~E., {Eriksson}, K.,
  \& {Gustafsson}, B. 2001, A\&A, 372, 885

\bibitem[{{Gonz{\'a}lez Delgado} {et~al.}(2003){Gonz{\'a}lez Delgado},
  {Olofsson}, {Schwarz}, {Eriksson}, {Gustafsson}, \&
  {Gledhill}}]{delgadoetal2003a}
{Gonz{\'a}lez Delgado}, D., {Olofsson}, H., {Schwarz}, H.~E., {et~al.} 2003,
  A\&A, 399, 1021

\bibitem[{{Groenewegen}(2012)}]{groenwegen2012}
{Groenewegen}, M.~A.~T. 2012, A\&A, 540, A32

\bibitem[{{Groenewegen} {et~al.}(2011){Groenewegen}, {Waelkens}, {Barlow},
  {Kerschbaum}, {Garcia-Lario}, {Cernicharo}, {Blommaert}, {Bouwman}, {Cohen},
  {Cox}, {Decin}, {Exter}, {Gear}, {Gomez}, {Hargrave}, {Henning},
  {Hutsem{\'e}kers}, {Ivison}, {Jorissen}, {Krause}, {Ladjal}, {Leeks}, {Lim},
  {Matsuura}, {Naz{\'e}}, {Olofsson}, {Ottensamer}, {Polehampton}, {Posch},
  {Rauw}, {Royer}, {Sibthorpe}, {Swinyard}, {Ueta}, {Vamvatira-Nakou},
  {Vandenbussche}, {van de Steene}, {van Eck}, {van Hoof}, {van Winckel},
  {Verdugo}, \& {Wesson}}]{groenewegenetal2011}
{Groenewegen}, M.~A.~T., {Waelkens}, C., {Barlow}, M.~J., {et~al.} 2011, A\&A,
  526, A162

\bibitem[{{G{\"u}sten} {et~al.}(2006){G{\"u}sten}, {Nyman}, {Schilke},
  {Menten}, {Cesarsky}, \& {Booth}}]{apexref}
{G{\"u}sten}, R., {Nyman}, L.~{\AA}., {Schilke}, P., {et~al.} 2006, A\&A, 454,
  L13

\bibitem[{{Hankins} {et~al.}(2018){Hankins}, {Herter}, {Maercker}, {Lau}, \&
  {Sloan}}]{hankinsetal2018}
{Hankins}, M.~J., {Herter}, T.~L., {Maercker}, M., {Lau}, R.~M., \& {Sloan},
  G.~C. 2018, ApJ, 852, 27

\bibitem[{{Herwig} \& {Austin}(2004)}]{herwigco2004}
{Herwig}, F. \& {Austin}, S.~M. 2004, ApJL, 613, L73

\bibitem[{{H{\"o}fner} \& {Andersen}(2007)}]{hofnerco2007}
{H{\"o}fner}, S. \& {Andersen}, A.~C. 2007, A\&A, 465, L39

\bibitem[{{Hony} \& {Bouwman}(2004)}]{honyco2004}
{Hony}, S. \& {Bouwman}, J. 2004, A\&A, 413, 981

\bibitem[{{Jager} {et~al.}(1998){Jager}, {Mutschke}, \&
  {Henning}}]{jageretal1998}
{Jager}, C., {Mutschke}, H., \& {Henning}, T. 1998, A\&A, 332, 291

\bibitem[{{Karakas} \& {Lattanzio}(2007)}]{karakasco2007}
{Karakas}, A. \& {Lattanzio}, J.~C. 2007, PASA, 24, 103

\bibitem[{{Kerschbaum} {et~al.}(2010){Kerschbaum}, {Ladjal}, {Ottensamer},
  {Groenewegen}, {Mecina}, {Blommaert}, {Baumann}, {Decin}, {Vandenbussche},
  {Waelkens}, {Posch}, {Huygen}, {De Meester}, {Regibo}, {Royer}, {Exter}, \&
  {Jean}}]{kerschbaumetal2010}
{Kerschbaum}, F., {Ladjal}, D., {Ottensamer}, R., {et~al.} 2010, A\&A, 518,
  L140

\bibitem[{{Kerschbaum} {et~al.}(2017){Kerschbaum}, {Maercker}, {Brunner},
  {Lindqvist}, {Olofsson}, {Mecina}, {De Beck}, {Groenewegen}, {Lagadec},
  {Mohamed}, {Paladini}, {Ramstedt}, {Vlemmings}, \&
  {Wittkowski}}]{kerschbaumetal2017}
{Kerschbaum}, F., {Maercker}, M., {Brunner}, M., {et~al.} 2017, A\&A, 605, A116

\bibitem[{{Kerschbaum} \& {Olofsson}(1999)}]{kerschbaumco1999}
{Kerschbaum}, F. \& {Olofsson}, H. 1999, A\&AS, 138, 299

\bibitem[{{Khouri} {et~al.}(2016){Khouri}, {Maercker}, {Waters}, {Vlemmings},
  {Kervella}, {de Koter}, {Ginski}, {De Beck}, {Decin}, {Min}, {Dominik},
  {O'Gorman}, {Schmid}, {Lombaert}, \& {Lagadec}}]{khourietal2016a}
{Khouri}, T., {Maercker}, M., {Waters}, L.~B.~F.~M., {et~al.} 2016, A\&A, 591,
  A70

\bibitem[{{Kirchschlager} \& {Wolf}(2014)}]{kirchschlageretal2014}
{Kirchschlager}, F. \& {Wolf}, S. 2014, A\&A, 568, A103

\bibitem[{{Knapp} {et~al.}(2003){Knapp}, {Pourbaix}, {Platais}, \&
  {Jorissen}}]{knappetal2003}
{Knapp}, G.~R., {Pourbaix}, D., {Platais}, I., \& {Jorissen}, A. 2003, A\&A,
  403, 993

\bibitem[{{Kov{\'a}cs}(2006)}]{kovacs2006}
{Kov{\'a}cs}, A. 2006, PhD thesis, Caltech

\bibitem[{{Kov{\'a}cs}(2008)}]{kovacs2008_crush}
{Kov{\'a}cs}, A. 2008, in Society of Photo-Optical Instrumentation Engineers
  (SPIE) Conference Series, Vol. 7020, Society of Photo-Optical Instrumentation
  Engineers (SPIE) Conference Series, 1

\bibitem[{{Maercker} {et~al.}(2012){Maercker}, {Mohamed}, {Vlemmings},
  {Ramstedt}, {Groenewegen}, {Humphreys}, {Kerschbaum}, {Lindqvist},
  {Olofsson}, {Paladini}, {Wittkowski}, {de Gregorio-Monsalvo}, \&
  {Nyman}}]{maerckeretal2012}
{Maercker}, M., {Mohamed}, S., {Vlemmings}, W.~H.~T., {et~al.} 2012, Nature,
  490, 232

\bibitem[{{Maercker} {et~al.}(2010){Maercker}, {Olofsson}, {Eriksson},
  {Gustafsson}, \& {Sch{\"o}ier}}]{maerckeretal2010}
{Maercker}, M., {Olofsson}, H., {Eriksson}, K., {Gustafsson}, B., \&
  {Sch{\"o}ier}, F.~L. 2010, A\&A, 511, A37+

\bibitem[{{Maercker} {et~al.}(2014){Maercker}, {Ramstedt}, {Leal-Ferreira},
  {Olofsson}, \& {Floren}}]{maerckeretal2014}
{Maercker}, M., {Ramstedt}, S., {Leal-Ferreira}, M.~L., {Olofsson}, G., \&
  {Floren}, H.~G. 2014, A\&A, 570, A101

\bibitem[{{Maercker} {et~al.}(2016){Maercker}, {Vlemmings}, {Brunner}, {De
  Beck}, {Humphreys}, {Kerschbaum}, {Lindqvist}, {Olofsson}, \&
  {Ramstedt}}]{maerckeretal2016}
{Maercker}, M., {Vlemmings}, W.~H.~T., {Brunner}, M., {et~al.} 2016, A\&A, 586,
  A5

\bibitem[{{Mancini} {et~al.}(2015){Mancini}, {Schneider}, {Graziani},
  {Valiante}, {Dayal}, {Maio}, {Ciardi}, \& {Hunt}}]{mancinietal2015}
{Mancini}, M., {Schneider}, R., {Graziani}, L., {et~al.} 2015, MNRAS, 451, L70

\bibitem[{{Matsuura} {et~al.}(2015){Matsuura}, {Dwek}, {Barlow}, {Babler},
  {Baes}, {Meixner}, {Cernicharo}, {Clayton}, {Dunne}, {Fransson}, {Fritz},
  {Gear}, {Gomez}, {Groenewegen}, {Indebetouw}, {Ivison}, {Jerkstrand},
  {Lebouteiller}, {Lim}, {Lundqvist}, {Pearson}, {Roman-Duval}, {Royer},
  {Staveley-Smith}, {Swinyard}, {van Hoof}, {van Loon}, {Verstappen}, {Wesson},
  {Zanardo}, {Blommaert}, {Decin}, {Reach}, {Sonneborn}, {Van de Steene}, \&
  {Yates}}]{matsuuraetal2015}
{Matsuura}, M., {Dwek}, E., {Barlow}, M.~J., {et~al.} 2015, ApJ, 800, 50

\bibitem[{{Mattsson} {et~al.}(2007){Mattsson}, {H{\"o}fner}, \&
  {Herwig}}]{mattssonetal2007}
{Mattsson}, L., {H{\"o}fner}, S., \& {Herwig}, F. 2007, A\&A, 470, 339

\bibitem[{{Micha{\l}owski} {et~al.}(2010){Micha{\l}owski}, {Watson}, \&
  {Hjorth}}]{michalowskietal2010}
{Micha{\l}owski}, M.~J., {Watson}, D., \& {Hjorth}, J. 2010, ApJ, 712, 942

\bibitem[{{Min} {et~al.}(2009){Min}, {Dullemond}, {Dominik}, {de Koter}, \&
  {Hovenier}}]{minetal2009}
{Min}, M., {Dullemond}, C.~P., {Dominik}, C., {de Koter}, A., \& {Hovenier},
  J.~W. 2009, A\&A, 497, 155

\bibitem[{{Olofsson} {et~al.}(1996){Olofsson}, {Bergman}, {Eriksson}, \&
  {Gustafsson}}]{olofssonetal1996}
{Olofsson}, H., {Bergman}, P., {Eriksson}, K., \& {Gustafsson}, B. 1996, A\&A,
  311, 587

\bibitem[{{Olofsson} {et~al.}(1993){Olofsson}, {Eriksson}, {Gustafsson}, \&
  {Carlstrom}}]{olofssonetal1993a}
{Olofsson}, H., {Eriksson}, K., {Gustafsson}, B., \& {Carlstrom}, U. 1993,
  ApJS, 87, 267

\bibitem[{{Olofsson} {et~al.}(2010){Olofsson}, {Maercker}, {Eriksson},
  {Gustafsson}, \& {Sch{\"o}ier}}]{olofssonetal2010}
{Olofsson}, H., {Maercker}, M., {Eriksson}, K., {Gustafsson}, B., \&
  {Sch{\"o}ier}, F. 2010, A\&A, 515, A27

\bibitem[{{Preibisch} {et~al.}(1993){Preibisch}, {Ossenkopf}, {Yorke}, \&
  {Henning}}]{preibischetal1993}
{Preibisch}, T., {Ossenkopf}, V., {Yorke}, H.~W., \& {Henning}, T. 1993, A\&A,
  279, 577

\bibitem[{{Ramstedt} {et~al.}(2011){Ramstedt}, {Maercker}, {Olofsson},
  {Olofsson}, \& {Sch{\"o}ier}}]{ramstedtetal2011}
{Ramstedt}, S., {Maercker}, M., {Olofsson}, G., {Olofsson}, H., \&
  {Sch{\"o}ier}, F.~L. 2011, A\&A, 531, A148

\bibitem[{{Ramstedt} {et~al.}(2008){Ramstedt}, {Sch{\"o}ier}, {Olofsson}, \&
  {Lundgren}}]{ramstedtetal2008}
{Ramstedt}, S., {Sch{\"o}ier}, F.~L., {Olofsson}, H., \& {Lundgren}, A.~A.
  2008, A\&A, 487, 645

\bibitem[{{Rau} {et~al.}(2017){Rau}, {Hron}, {Paladini}, {Aringer}, {Eriksson},
  {Marigo}, {Nowotny}, \& {Grellmann}}]{rauetal2017}
{Rau}, G., {Hron}, J., {Paladini}, C., {et~al.} 2017, A\&A, 600, A92

\bibitem[{{Rouleau} \& {Martin}(1991)}]{rouleauco1991}
{Rouleau}, F. \& {Martin}, P.~G. 1991, ApJ, 377, 526

\bibitem[{{Schneider} {et~al.}(2014){Schneider}, {Valiante}, {Ventura},
  {dell'Agli}, {Di Criscienzo}, {Hirashita}, \& {Kemper}}]{schneideretal2014}
{Schneider}, R., {Valiante}, R., {Ventura}, P., {et~al.} 2014, MNRAS, 442, 1440

\bibitem[{{Sch{\"o}ier} {et~al.}(2005){Sch{\"o}ier}, {Lindqvist}, \&
  {Olofsson}}]{schoieretal2005}
{Sch{\"o}ier}, F.~L., {Lindqvist}, M., \& {Olofsson}, H. 2005, A\&A, 436, 633

\bibitem[{{Sch{\"o}ier} \& {Olofsson}(2001)}]{schoierco2001}
{Sch{\"o}ier}, F.~L. \& {Olofsson}, H. 2001, A\&A, 368, 969

\bibitem[{{Siringo} {et~al.}(2009){Siringo}, {Kreysa}, {Kov{\'a}cs},
  {Schuller}, {Wei{\ss}}, {Esch}, {Gem{\"u}nd}, {Jethava}, {Lundershausen},
  {Colin}, {G{\"u}sten}, {Menten}, {Beelen}, {Bertoldi}, {Beeman}, \&
  {Haller}}]{labocaref}
{Siringo}, G., {Kreysa}, E., {Kov{\'a}cs}, A., {et~al.} 2009, A\&A, 497, 945

\bibitem[{{Steffen} \& {Sch{\"o}nberner}(2000)}]{steffenco2000}
{Steffen}, M. \& {Sch{\"o}nberner}, D. 2000, A\&A, 357, 180

\bibitem[{{Suh}(2000)}]{suh2000}
{Suh}, K.-W. 2000, MNRAS, 315, 740

\bibitem[{{Vlemmings} {et~al.}(2017){Vlemmings}, {Khouri}, {O'Gorman}, {De
  Beck}, {Humphreys}, {Lankhaar}, {Maercker}, {Olofsson}, {Ramstedt}, {Tafoya},
  \& {Takigawa}}]{vlemmingsetal2017}
{Vlemmings}, W., {Khouri}, T., {O'Gorman}, E., {et~al.} 2017, Nature Astronomy,
  1, 848

\bibitem[{{Watson} {et~al.}(2015){Watson}, {Christensen}, {Knudsen}, {Richard},
  {Gallazzi}, \& {Micha{\l}owski}}]{watsonetal2015}
{Watson}, D., {Christensen}, L., {Knudsen}, K.~K., {et~al.} 2015, Nature, 519,
  327

\bibitem[{{Woitke}(2006)}]{woitke2006}
{Woitke}, P. 2006, A\&A, 460, L9

\bibitem[{{Xu} {et~al.}(2016){Xu}, {Lin}, {Zhang}, \& {Hao}}]{xuetal2016}
{Xu}, Y., {Lin}, Y., {Zhang}, J., \& {Hao}, J. 2016, ApJ, 825, 111

\bibitem[{{Zubko} {et~al.}(1996){Zubko}, {Mennella}, {Colangeli}, \&
  {Bussoletti}}]{zubkoetal1996}
{Zubko}, V.~G., {Mennella}, V., {Colangeli}, L., \& {Bussoletti}, E. 1996,
  MNRAS, 282, 1321

\end{thebibliography}

\end{document}